\documentclass[journal]{IEEEtran}
\usepackage{amsmath,amsfonts}
\usepackage{algorithmic}
\usepackage{array}
\usepackage[caption=false,font=normalsize,labelfont=sf,textfont=sf]{subfig}
\usepackage{textcomp}
\usepackage{stfloats}
\usepackage{url}
\usepackage{verbatim}
\usepackage{graphicx}
\usepackage{cite}
\usepackage{xcolor}
\usepackage{orcidlink}
\usepackage{tabularray}
\usepackage{pifont}

\usepackage{multirow}
\usepackage[ruled,vlined]{algorithm2e}

\usepackage{tabularx}
\usepackage{xparse}
\usepackage{makecell}
\usepackage{microtype}
\microtypesetup{activate=false}

\begin{document}

\title{NITRO: High-Performance 3D NAND Flash-Based In-Storage Computing with Enhanced Activation Dataflow}

\author{{\textls[-25]{Sanghun Shin\raisebox{0.5ex}{\orcidlink{0009-0004-7708-3737}}\textsuperscript{\textdagger,*},
        Sangyeon Kim\raisebox{0.5ex}{\orcidlink{0009-0009-5846-1508}}\textsuperscript{\textdagger,*},~\textit{Graduate Student Member,~IEEE,}
        Gisan Ji\raisebox{0.5ex}{\orcidlink{0009-0009-0061-433X}}\textsuperscript{\textdagger},
        and~Sungju Ryu\raisebox{0.5ex}{\orcidlink{0000-0002-0254-391X}}\textsuperscript{\textdaggerdbl,\S},~\textit{Member,~IEEE}}
        \\
        \textsuperscript{\textdagger}Department of Electronic Engineering, Sogang University, Seoul, Republic of Korea
        \\
        \textsuperscript{\textdaggerdbl}Department of System Semiconductor Engineering, Sogang University, Seoul, Republic of Korea
        \\
        \textsuperscript{\textdagger}\{sanghun, sangyeonkim, gisanji\}@sogang.ac.kr, \textsuperscript{\textdaggerdbl}sungju@sogang.ac.kr}


\thanks{This work is an extended version of a paper accepted for publication in the Proceedings of the DATE 2026 \cite{shin2026nitro}. Compared to the conference version, this manuscript provides a more comprehensive description of the technical background, detailed explanations of the proposed architecture, and an extended evaluation including throughput results, an additional S-FLASH baseline, and area analysis. Details are summarized in Appendix \ref{appendices}.}

\thanks{\textsuperscript{*}Equal contribution.}
\thanks{\textsuperscript{\S}Corresponding Author.}
}

\markboth{}{}

\maketitle
\begin{abstract}


In-storage computing (ISC) is considered a next-generation memory architecture for its ability to relieve the data bottleneck between the host and the memory. 
While the required resources of large language models (LLMs) have increased significantly in recent years, the memory density has not scaled accordingly. 
\textcolor{black}{
Recently, several works have studied NAND flash-based processing-in-memory (NAND-PIM) schemes to exploit the high density of the memory.
However, they do not address the dataflow/buffer for the intermediate values, so a simple method is to deal with the values in the slow flash memory array.
To overcome such a limitation, 
we propose a high-performance NAND flash-based ISC architecture with enhanced activation buffering.
Instead of using the very slow flash memory array for the intermediate values, our architecture buffers the values in a fast DRAM subsystem.
This approach effectively handles the high-latency penalties when activations are programmed into slower TLC NAND flash.
We also introduce a distributed dataflow approach for the NAND-PIM array.
This approach maximizes computational parallelism by employing efficient intra-plane data mapping.
The results show that our proposed architecture achieves significant performance improvements, reducing the inference latency by up to 99.7\% compared to the baseline.
}

\end{abstract}


\section{Introduction}
\label{section_intro}

Transformer-based LLMs have revolutionized a wide spectrum of AI applications from language understanding to code synthesis within just a few years.
The rapid scaling of model parameters enabled these models to learn from vast datasets and show impressive capabilities \cite{zhang2022opt}.
However, as the model parameters grow exponentially, the memory resources and memory bandwidth play a critical role on LLM performance. 
In particular, frequent data movements between processor and memory lead to a significant memory bottleneck, resulting in increased inference latency and underutilization of powerful compute resources.

To handle the memory bottleneck problem, PIM has been considered a promising solution.
The PIM architecture integrates the computation logic inside the memory units, enabling more efficient bandwidth utilization and reduced latency.

While previous DRAM, SRAM, and ReRAM-based PIM approaches offer benefits \cite{lee2021hardware, liu2023hardsea}, the large size of LLMs necessitates exploring memory technologies with better density and scalability.
3D NAND flash-based SSD emerges as a potential candidate thanks to its exceptional storage density, non-volatility, and low cost per bit.
This has motivated exploration of NAND flash-based computation through two main paradigms: in-storage computing (ISC) and processing-in-memory (PIM).
Modern SSDs enable ISC through their high internal bandwidth between the controller and concurrently operated flash chips, or by inserting specific processors.
Several recent works have applied ISC to parts of the DNN workflow.
Cambricon-LLM \cite{yu2024cambricon} introduced a hybrid architecture by connecting the NPU directly to the NAND flash chip, leveraging flash for weight storage and on-die processing tasks.
However, these ISC-based approaches often rely on SSD controller resources or, as in Cambricon-LLM, require an external processing unit for acceleration.

Alternatively, NAND-PIM performs computations directly inside the NAND flash memory array.
The analog current-sum method has been widely used to perform matrix-vector multiplication (MVM) by leveraging the crossbar array structure of the NAND array \cite{lue2019optimal,lee20223d, kang2021s}.
For example, 3D-FPIM \cite{lee20223d} designed a quantization-aware conversion and low-power decoder to mitigate accuracy loss and WL energy overhead, respectively.
S-Flash \cite{kang2021s} optimized energy efficiency by exploiting bit-level sparsity in activations and weights, allowing more current accumulation within analog-to-digital converters (ADC) level constraints.
While these analog PIM methods demonstrate potential for efficient MVM execution, they often focus primarily on the MVM kernel itself.
They also employ simplified designs (e.g., small array sizes, single-level cell (SLC) type) instead of real-world commercial flash memory configuration and lack a comprehensive strategy for managing the complex dataflows required for LLM inference.
Therefore, efficiently handling the large size of intermediate activation data generated between MVM computations remains a challenge unaddressed by prior analog NAND-PIM works.

In this work, we propose NITRO: a novel 3D NAND flash-based ISC architecture with analog PIM methodology for Transformer LLM acceleration.
NITRO overcomes the limitations of previous approaches by introducing an optimized dataflow strategy that efficiently manages intermediate results inside the SSD subsystem.

Our contributions can be summarized as follows:

\begin{enumerate}\itemsep3pt

\item We analyze the workflow of LLMs and identify the possibilities of performance improvements over the previous NAND-PIM by buffering intermediate activation data in DRAM to reduce the read/write latency when using slow flash memory array.

\item We present the optimized dataflow scheme to leverage the multi-plane parallelism of NAND flash dies.


\item Based on the contributions above, we propose a NITRO ISC architecture.

\end{enumerate}

The following sections are organized as follows. 
Section \ref{section_prelim} briefly explains the background of 3D NAND-PIM. 
Section \ref{section_proposed} introduces the proposed methodology.
Section \ref{section_result} describes the experimental setup and presents the results. 
Finally, Section \ref{section_conclusion} concludes the paper.


\section{Preliminaries}
\label{section_prelim}


\subsection{3D NAND Flash}
\label{section_prelim_3d_nand}

\begin{figure}[t]
\vspace{-5mm}
\centering
\includegraphics[width=0.48\textwidth]{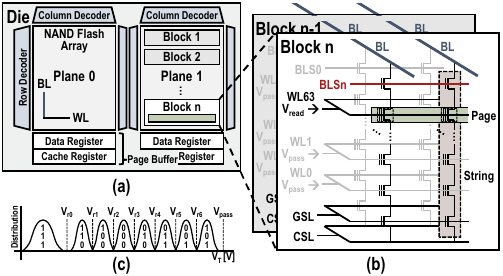}
\vspace{-2mm}
\caption{Illustration of 3D NAND flash memory: (a) Die-level and (b) plane-level architectures.
(c) Voltage distribution of TLC.}
\label{prelim_3D_NAND_Arch}
\end{figure}

Scalability and cost-effectiveness have made the 3D NAND flash a widely used memory architecture in high-density storage devices.
Fig. \ref{prelim_3D_NAND_Arch}a shows its die-level hierarchy.
Multiple independent planes (often 2-4) form a single die, where each plane is equipped with its own NAND flash array, row/column decoders, and data registers, enabling concurrent access across planes.
Each plane is partitioned into hundreds of blocks, each block consisting of numerous NAND strings (Fig. \ref{prelim_3D_NAND_Arch}b).
Each string comprises a vertical stack of memory cells connected in series between the bitline selector (BLS) transistor and the ground select line (GSL) transistor.
Strings in the same column share a bitline (BL), while cells at the same vertical level across multiple strings share a wordline (WL).
A group of cells connected to the same WL forms a page, the fundamental unit for read and program operations, often ranging from 4 to 16 KB in size.
To maximize bit density, modern 3D NAND devices employ a multi-level cell technology, such as triple-level cell (TLC) using the charge trap flash (CTF) cell structure \cite{tehrani2013advancement}.
Unlike SLC, which stores one bit using two different threshold voltage states ($V_{T}$), TLC cells can hold three bits per cell.
It is operated precisely by programming and distinguishing among eight distinct $V_{T}$ levels inside the cell's operating window (Fig. \ref{prelim_3D_NAND_Arch}c).

Reading data from the TLC NAND flash block involves a sequence of voltage operations.
First, the target block with specific page is selected by activating its BLS and GSL.
The row decoder applies the read reference voltage ($V_{read}$) to the target WL.
Concurrently, unselected WLs in the same block receive a pass voltage ($V_{pass}$), higher than the highest programmed $V_{T}$ state (Fig. \ref{prelim_3D_NAND_Arch}c).
The $V_{pass}$ ensures that unselected cells act as a pass transistor regardless of their $V_{T}$ level.
For the target cell in a selected page, the $V_{T}$ level determines whether it acts as a pass transistor or turns off at the applied $V_{read}$. 
The current determined by the target cell in the selected string flows through the connected BL, where it discharges the load capacitor located inside the sense amplifier \cite{park2022flash}.
Then it is decoded by the sense amplifier logic.
Multiple read operations are required to identify the three bits stored in a TLC cell.
The row decoder sequentially applies different $V_{read}$, corresponding to the reference voltages ($V_{r0}$ - $V_{r6}$) shown in Fig. \ref{prelim_3D_NAND_Arch}c.
By identifying the cell's conductance at different $V_{read}$ levels, the sense amplifier and latching circuit decode the specific $V_{T}$ state and its corresponding 3-bit data ('111' to '000').

\begin{figure}[t]
\vspace{-5mm}
\centering
\includegraphics[width=0.48\textwidth]{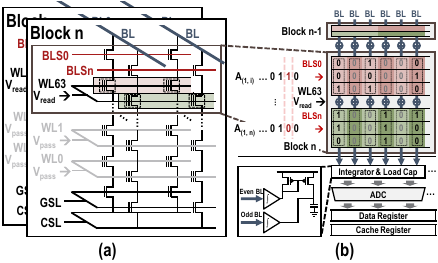}
\vspace{-2mm}
\caption{
The matrix-vector multiplication mechanism in 3D NAND flash memory.
}
\vspace{-5mm}
\label{prelim_3D_NAND_MVM}
\end{figure}
\begin{figure*}[t]
\vspace{-5mm}
\centering
\includegraphics[width=0.98\textwidth]{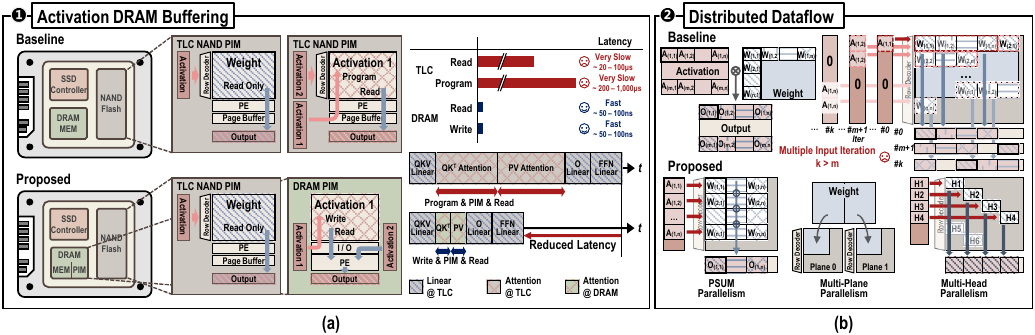}
\vspace{-2mm}
\caption{
Overview of our contribution: (a) Activation DRAM buffering and (b) distributed dataflow.
}
\vspace{-5mm}
\label{overview_arch}
\end{figure*}

\subsection{3D NAND Flash-Based MVM}
\label{section_prelim_MVM}

Previous studies \cite{lue2019optimal, lee2020neuromorphic,lee20223d, kang2021s} have implemented MVM operations on 3D NAND flash memory arrays (Fig. \ref{prelim_3D_NAND_MVM}a).
Weights ($W$) are stored in memory cells by programming their corresponding $V_T$ levels, where each level dictates cell's conductance \cite{lee2020neuromorphic}.
Activation ($A$) values, encoded as bit-wise voltage pulses, are applied to specific NAND strings via their BLSs.
Fig. \ref{prelim_3D_NAND_MVM}b illustrates MVM operation ($O=A \times W$) inside the NAND array.
To compute a dot product, a WL containing target weights is selected and biased with appropriate $V_{read}$.
$V_{pass}$ is applied to unselected WLs, turning their cells into pass transistors.
Based on the $A$ data bits, corresponding BLSs are activated.
This connects each associated NAND string to its respective BL.
Current through the activated NAND string, proportional to the conductance of the selected cell, performs analog multiplication between the input and the weight.
Accumulated current from NAND strings onto a connected BL generates an analog value representing an output ($O$) value.
For signed weight matrices, paired BLs are used \cite{lee2020neuromorphic}.
Positive weights are programmed into cells on even BLs, and negative weight's magnitudes into cells on odd BLs.
During MVM, both even and odd BLs independently accumulate current from their respective activated cells.
The current from even BL charges a load capacitor, while the current from odd BL discharges it.
The resulting capacitor voltage is then digitized by ADCs.
This analog-current summation method leverages 3D NAND flash array's massive parallelism to concurrently process multiple multiply-accumulate (MAC) operations.


\section{Proposed Architecture}
\label{section_proposed}


\subsection{Overview}
\label{section_proposed_overview}

This section details the proposed NITRO architecture including heterogeneous PIM capabilities inside the SSD to efficiently process Transformer-based models. 
Fig. \ref{overview_arch} provides an overview, comparing the baseline with our approach.
Executing Transformer-based models solely on conventional NAND-PIM is challenging due to the slow latency characteristics of NAND flash.
As illustrated in the baseline scenario of Fig. \ref{overview_arch}a, if generated intermediate activations programming into NAND array for subsequent processing, the slow program latency of the NAND flash creates a bottleneck.
Consequently, this latency negatively impacts overall inference speed. 
Meanwhile, DRAM has a latency of read and write several orders of magnitude faster than the NAND flash \cite{sudan2012nand}.
But, previous NAND-PIMs typically focus on MVM using NAND flash array only.
To improve the workflow, we integrate processing capabilities into both the NAND and DRAM. 
In this case, we store the weights in the NAND flash array and perform the MVM in the same way as conventional NAND-PIM architectures.
Considering that the write operation of the NAND flash is much slower than the read operation, writing output values in the NAND may significantly degrade the throughput.
Therefore, we select the DRAM as the storage of the activations instead of the NAND flash array to increase the overall inference speed.
Afterwards, matrix computations between activation values at the attention layers are performed by adopting the conventional DRAM-PIM approach.
We offload specific Transformer blocks, particularly those involving frequent read/write operations on intermediate activation data, to DRAM-PIM.
This process significantly reduces the overall inference latency compared to the NAND-only approach.

Furthermore, we also explain the additional limitations of previous works \cite{lee20223d, kang2021s} in utilizing the NAND-PIM array.
These earlier methods often targeted simpler DNN, showing inefficiency when the dimensions of mapping weights do not align well with the large physical dimensions of NAND flash arrays.
Consequently, naive mappings can lead to suboptimal resource usage.
As illustrated in the baseline of Fig. \ref{overview_arch}b, such approaches often require multiple sequential iterations for the computation of even a single row of an output matrix, thereby increasing the latency of MVM.
To maximize the computational throughput of the NAND-PIM, NITRO enhances the data mapping scheme inside the NAND flash array.
This includes leveraging parallel accumulation of partitioned array during linear block computations and enabling concurrent processing of multiple heads in attention blocks.
Additionally, NITRO distributes weight data across the NAND-PIM planes to effectively exploit multi-plane parallelism to increase throughput.
By combining latency reduction through DRAM-PIM and throughput enhancement by optimized dataflow in NAND flash array, NITRO reduces the inference latency inside the storage devices.
\begin{figure*}[t]
\vspace{-5mm}
\centering
\includegraphics[width=0.98\textwidth]{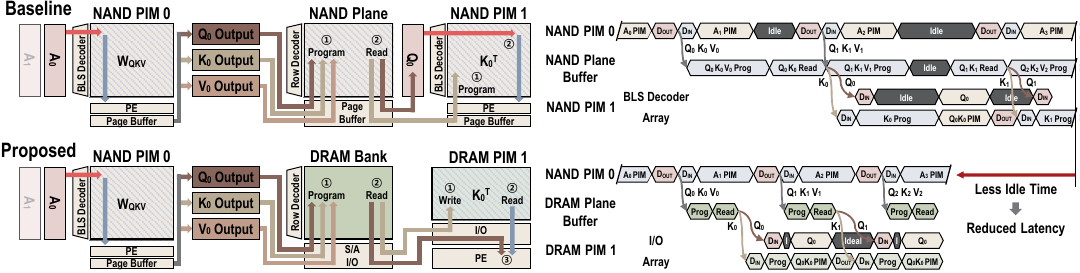}
\caption{
Comparison between the baseline and the proposed activation DRAM buffering.
}
\vspace{-5mm}
\label{contribution_1}
\end{figure*}

\begin{figure}[t]
\centering
\includegraphics[width=0.48\textwidth]{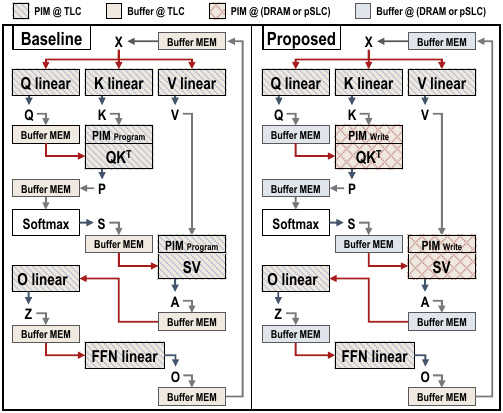}
\caption{
Activation DRAM buffering using heterogeneous TLC and DRAM memory types for various Transformer layers.
}
\vspace{-5mm}
\label{contribution_1_dataflow}
\end{figure}

\subsection{Activation DRAM buffering}
\label{section_proposed_AB}

Executing entire Transformer blocks only in the TLC NAND-PIM can create a huge bottleneck, mostly from the attention blocks.
These blocks often require the multiplication of two intermediate activation matrices (e.g., $QK^T$).
To process this multiplication using NAND-PIM, one activation matrix ($K^T$) must first be programmed into NAND cells.
Since TLC NAND exhibits a few hundred microseconds of program time, a critical performance bottleneck is created.
This could disrupt the pipeline process of Transformer blocks and increase the idle time of other NAND-PIM.
For instance, after generating $Q_0$ and $K_0$ from $A_0$, the system must wait for programming $K_0^T$ before $Q_0K_0^T$ computation can begin (Fig. \ref{contribution_1}).
Further processing of $A_1$ is also stalled due to resource contention, increasing the overall inference latency.
To address this critical latency issue, NITRO adopts a hybrid PIM architecture, offloading intermediate activation data to DRAM and attention block computation to DRAM-PIM. 
Intermediate activations such as $K_0$ generated from a preceding layer are rapidly written to the DRAM buffer.
These activations are then transposed to be written in DRAM-PIM with low latency.
The attention computation of $Q_0K_0^T$ can also achieve low latency from the fast DRAM side, since $K_0^T$ is read from the DRAM array to be computed with $Q_0$ inside the processing unit.
This could effectively reduce the latency compared to the NAND programming steps.

As shown in Fig. \ref{contribution_1_dataflow}, based on the data access characteristics, we map the various computational blocks of Transformer layers to the most suitable PIM subsystem.
Linear blocks involve large matrix-matrix multiplication (MMM) between input activations and trained static weights.
Therefore, we assign them to the TLC NAND-PIM array.
No programming of weight data is required during inference, and the high density of TLC NAND flash is suitable for storing large weight matrices.
Conversely, attention blocks heavily rely on activation-activation multiplications, which suffer from the programming bottleneck in NAND flash.
To mitigate the programming latency, they are offloaded to DRAM-PIM instead of NAND-PIM.
Between the Transformer blocks, the intermediate activation data is stored inside the DRAM buffer to be processed in a subsequent block.

For DRAM-less SSD configuration, the buffering of intermediate data and attention computation can instead be mapped to NAND blocks operating in a faster pseudo-SLC (pSLC) mode.
However, while pSLC programming is faster than TLC, it remains slower than DRAM, thereby creating unavoidable latency during attention computations.
We performed the experiments for both pSLC and DRAM cases, which will be discussed in the later Section \ref{section_result}.

\begin{figure*}[t]
\vspace{-5mm}
\centering
\includegraphics[width=0.98\textwidth]{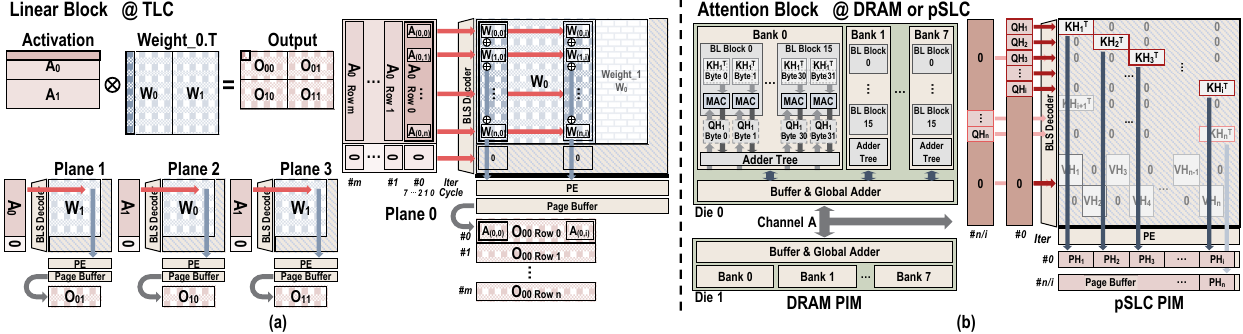}
\vspace{-2mm}
\caption{Distributed dataflow for (a) linear blocks and (b) attention blocks.}
\vspace{-5mm}
\label{contribution_2}
\end{figure*}

\subsection{Distributed Dataflow}
\label{section_proposed_DD}
\subsubsection{Linear Blocks}
For Transformer-based models to process MVM between activation and weights effectively on NAND array, we first placed weights on a single WL layer without modifying the footprint of the tensors on the array.
For instance, a single query weight matrix in LLaMa-2 13B requires approximately 5120 rows and 15360 columns (described in Section \ref{section_proposed_NITRO_Arch}), which can be placed in the layer of 3D NAND flash array.
To maximize throughput using the plane-level parallelism of 3D NAND flash, we employ several data partitioning strategies.
First, the weight matrix ($W$) is vertically partitioned into $W_0$ and $W_1$ after the transpose operation and distributed across multiple planes as illustrated in Fig. \ref{contribution_2}a.
This arrangement matches the NAND-PIM architecture, where current accumulation occurs along the BLs.
We can also consider horizontal partitioning, but it
would underutilize resources, as blocks without relevant weight data would remain idle, and additional operations to sum partial outputs is required.
Hence, vertical partitioning is optimal when the weight matrix's row dimension fits inside the NAND array's WL dimension.
It allows each output row to be computed via MVM inside the planes, with simple concatenation of partial output matrices ($O_{00}$, $O_{01}$, $O_{10}$, $O_{11}$).
If the size of the weight matrix is larger than the row size of the NAND array, horizontal partitioning is needed to maximize utilization of the array resource.

Second, during the prefill stage of inference, the activation matrices contain multiple rows corresponding to the input tokens.
We also have partitioned $A$ horizontally into $A_0$ and $A_1$. As shown in Fig. \ref{contribution_2}a, $A_0$ is multiplied with $W_0$ and $W_1$ in planes 0 and 1, respectively.
$A_1$ is concurrently multiplied with $W_0$ and $W_1$ in planes 2 and 3, respectively. 
This allows parallel processing of tokens, effectively reducing the number of MVM iterations and latency compared to processing $A$ row-by-row.
During the decode stage, where $A$ is often a single vector, this horizontal partitioning is not applied. 
However, the input vector is broadcast to planes 0 and 1 to generate partial outputs $O_{00}$ and $O_{01}$, respectively.
This multi-plane parallel computation significantly accelerates linear block execution compared to non-parallel approaches.

\subsubsection{Attention Blocks}
Attention blocks compute relevance scores via multi-head attention \cite{vaswani2017attention}. 
The implementation depends on the target hardware.
In systems with DRAM-PIM, as shown in Fig. \ref{contribution_2}b (left), the multiple banks inside the DRAM can be operated in parallel.
Different attention heads are distributed across separate DRAM banks. 
Each bank's process unit computes the head-wise attention operation (e.g., $QK^T$) in a vector-by-vector manner,
leveraging the bank-level parallelism.

In the case of DRAM-less SSDs, multi-head attention operations are mapped to the NAND array operating in a faster pSLC mode.
Fig. \ref{contribution_2}b (right) illustrates the mapping strategy for computing $QK^T$ and $SV$ in pSLC NAND flash array.
Multi-head key matrices ($KH_1^T$, $KH_2^T$, ..., $KH_n^T$) are distributed across different BL groups using a diagonal placement scheme.
Multi-head query matrices ($QH_1$, $QH_2$, ..., $QH_n$) are input row-by-row via the BLS decoder and subsequently multiplied with their relevant key matrices.
Only a subset of heads (e.g., heads $\textit{0}-\textit{i}$) can be processed in a single iteration.
This is because if the current on a BL is accumulated from two or more different heads, the output cannot be guaranteed.
Meanwhile, $score \cdot V$ ($SV$) computation requires different mapping challenges compared to the $QK^T$ computation.
Typically, in Transformer-based models, the token size is much larger than the vector size of each attention head.
In the $QK^T$ computation, the dot product operations are performed in the vector-size-direction, so a single dot product chunk is small.
As a result, thanks to the small chunk size, we can simultaneously deal with a large number of attention heads in a NAND-PIM array.
On the other hand, in the $SV$ computation, the dot products are computed in the token-size-direction.
Therefore, the number of value heads ($VH$) that can be processed concurrently in a single iteration is limited (e.g., 2 heads).
This requires more iterations to complete the full attention output calculation.

\begin{figure}[t]
\centering
\includegraphics[width=0.48\textwidth]{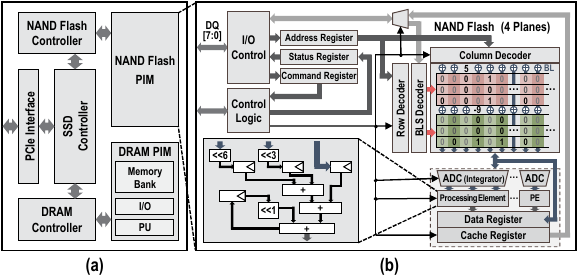}
\vspace{-2mm}
\caption{Top-level architecture of NITRO.}
\vspace{-5mm}
\label{NITRO_Architecture}
\end{figure}

\begin{figure*}[t]
\vspace{-5mm}
\centering
\includegraphics[width=0.98\textwidth]{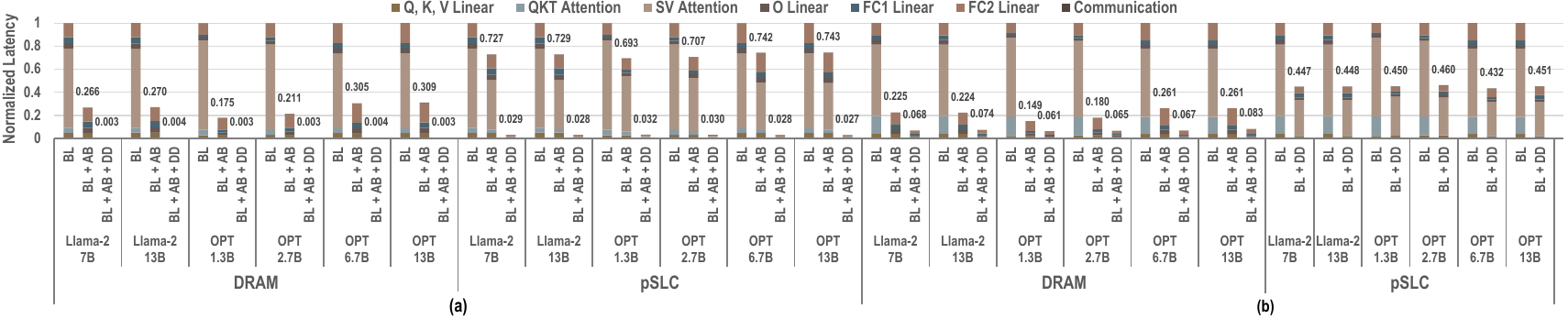}
\vspace{-2mm}
\caption{\textcolor{black}{Normalized inference latency of (a) linear at TLC, attention at DRAM/pSLC with baseline as 3D-FPIM, (b) linear at TLC, attention at DRAM/pSLC with baseline as S-FLASH. BL: Baseline. AB: Activation buffering. DD: Distributed dataflow.}}
\vspace{-5mm}
\label{Fig_latency}
\end{figure*}
\begin{figure}[t]
\centering
\includegraphics[width=0.48\textwidth]{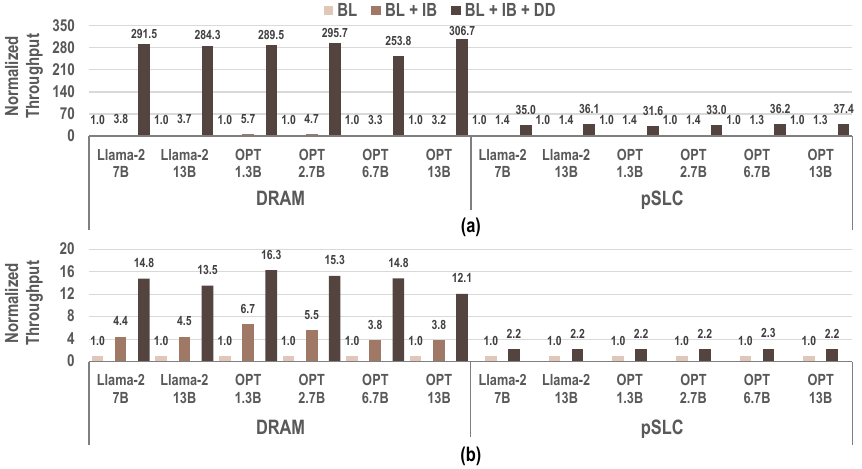}
\vspace{-2mm}
\caption{\textcolor{black}{Normalized inference throughput of (a) linear at TLC, attention at DRAM/pSLC with baseline as 3D-FPIM, (b) linear at TLC, attention at DRAM/pSLC with baseline as S-FLASH. BL: Baseline. AB: Activation buffering. DD: Distributed dataflow.}}
\vspace{-2mm}
\label{Fig_Throughput}
\end{figure}

\subsection{NITRO Architecture}
\label{section_proposed_NITRO_Arch}

Fig. \ref{NITRO_Architecture} illustrates the overall NITRO architecture, capable of ISC inside an SSD.
Fig. \ref{NITRO_Architecture}a shows the internal architecture, similar to a conventional SSD.
The architecture includes a NAND flash subsystem and a DRAM subsystem managed by an SSD controller.
In NITRO, the DRAM subsystem manages multiple functions from storing logical-to-physical mapping tables to buffering intermediate data and also executing computation tasks like attention block or activation functions using dedicated PIM banks.
\textcolor{black}{
To support these functionalities, we adopted a combination of the McDRAM \cite{shin2018mcdram}, an LPDDR4-based PIM developed by a DRAM manufacturer, and an LPDDR4 SDRAM \cite{micronlpddr4}, which is a DRAM type commonly used inside SSD \cite{sutardja20151}.
}
The detailed explanation of NAND-PIM is shown in Fig. \ref{NITRO_Architecture}b.
NAND flash die contains four independent planes, each equipped with standard components such as row/column decoders and data/cache registers.
NITRO also integrates specialized hardware to enable PIM computations.
\textcolor{black}{
This includes a BLS decoder for NAND string activation based on input data, ADCs equipped with integrators and capacitors for digitizing analog computation outputs, and processing elements (PE) for digitally accumulating PIM results.
Additionally, to support conventional NAND flash read/program operations, a bypass path directly connecting the BLs to the data register is also included.
}

A simple explanation of computation with NITRO architecture for 8-bit weights is as follows.
Before execution of LLM inference, weight matrices quantized to INT8 format are programmed into TLC NAND flash cell array.
NITRO employs a sign-and-magnitude representation for weights.
The 7-bit magnitude of each data value is first padded with two leading zeros to form a 9-bit representation.
This 9-bit value is then split into three parts and stored across multiple physical NAND cells in a page-wise direction.
Positive weight data are programmed into cells on even BLs, while negative weight data are stored on corresponding odd BLs.
During MVM operation, the input activation vector, often represented in 2's complement format, is processed bit-serially.
The BLS decoders activate the NAND string according to the input bit and selectively connect NAND strings to BLs.
The current proportional to the selected cell's conductance flows through the activated NAND string.
Analog computation occurs via differential sensing using paired BLs.
Current from the even BL charges an integrator's load capacitor, while current from the odd BL discharges it.
This differential voltage is then digitized by ADCs.
The digitized partial result, corresponding to the MSB, CSB, and LSB weights, are then fed into processing elements (PE).
The PE performs two key operations.
First, for each input bit, it reconstructs the full partial product by shifting and summing the digitized components.
Second, it accumulates these partial products across all bits of the input vector.
To achieve this, the accumulated sum from the previous input bit is added to the current bit's partial product.
This intermediate sum is then shifted by one bit to align with the next input bit's partial product.
The final MVM result, potentially up to 16-bits, is wider than a conventional output.
Therefore, it is split into two 8-bit portions and stored across the data and cache registers.
This output is then ready for transfer via the I/O interface.
Overall, the NITRO architecture delivers a PIM-compatible ISC SSD solution by integrating computational logic into both the DRAM and NAND flash die, enabling the acceleration of data-intensive workloads inside the storage system.


\section{Experimental Results}
\label{section_result}


\subsection{Experimental Setup}
\label{section_result_setup}


\textbf{Circuit-level Simulation:}
To calculate the latency of NITRO architecture, a NVSIM\cite{dong2012nvsim}-based modified simulator 3D-FPIM \cite{lee20223d} was used. 
The simulator supports emulation of the behavior of a 3D NAND flash die and 3D NAND-PIM.
The timing parameters and hardware configurations for TLC and pSLC mode NAND flash were based on 256Gb TLC devices \cite{ymtc256}.
The digital circuit of the processing element for NAND-PIM was synthesized using Synopsys Design Compiler in 28nm node, and then it was scaled to 7nm using \cite{stillmaker2017scaling} to reflect recent technology node.
Parameters related to DRAM-PIM were adopted from existing LPDDR4 DRAM-PIM like McDRAM \cite{shin2018mcdram} and specifications of Micron's LPDDR4 SDRAM \cite{micronlpddr4}.
Using these circuit-level and architectural parameters, we developed in-house simulator. 

\textbf{Baseline:}
We evaluated the NITRO architecture against two representative NAND-PIM baseline approaches that use the analog current-sum method: (i) 3D-FPIM \cite{lee20223d} and (ii) S-FLASH \cite{kang2021s}.
For the 3D-FPIM baseline, large weight matrices of the Transformer models are partitioned into fixed-size 128 $\times$ 128 sub-matrices.
Where partition size of 512 $\times$ 128 is used for S-FLASH baseline, following the specification of each prior work.
Note that previous works did not considered the case of placing weight matrices into NAND flash arrays larger than the weight matrices.
In both baseline models, these sub-matrices are sequentially programmed onto the rows (WLs) of a conventional NAND flash array (with a 16KB page size), row-by-row until the page is filled. 
The sequential tiling process is conceptually illustrated in the baseline dataflow of Fig. \ref{overview_arch}b.
Furthermore, S-FLASH operates under the constraint of using low-resolution ADC-like sense amplifiers (SAs).
The limited resolution restricts the number of NAND strings that participate in the MVM operation without exceeding the SA's sensing range or precision.
This limitation often leads to additional processing iterations, even within a single weight tile, as only subset of the MAC operation can be performed per ADC cycle.
\textcolor{black}{
We have analyzed three cases: the baseline approach ($BL$), NITRO with activation buffering in DRAM or pSLC ($BL+AB$), and fully optimized NITRO architecture with both activation buffering and distributed dataflow in NAND-PIM ($BL+AB+DD$).
For the $BL+AB$ and $BL+AB+DD$ cases, we considered both DRAM and pSLC for the activation buffering and attention block computation.
Linear blocks are processed in TLC mode of NAND-PIM in both scenarios.
Note that the $BL+AB$ configuration is omitted for S-FLASH baseline when attention block is mapped to pSLC, as S-FLASH operates on the SLC NAND-PIM.
}

\begin{figure*}[t]
\vspace{-5mm}
\centering
\includegraphics[width=0.98\textwidth]{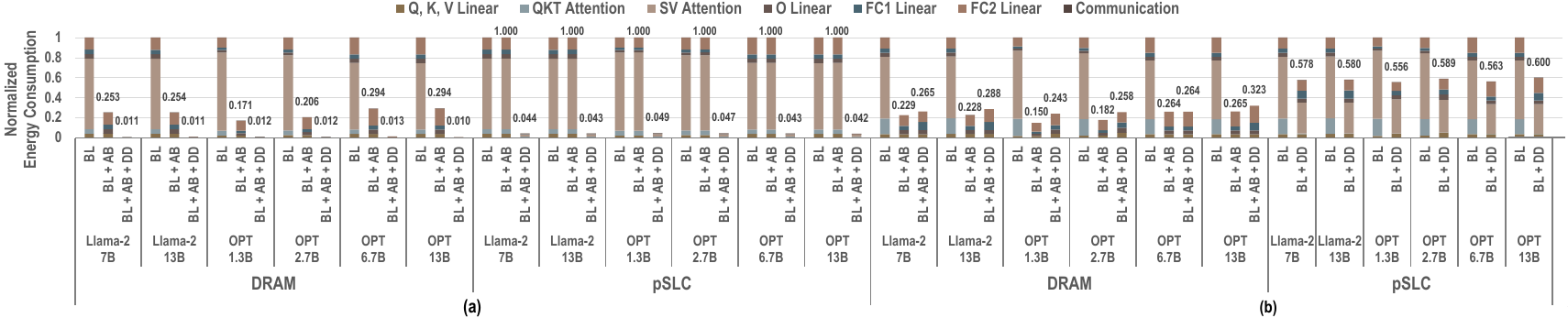}
\vspace{-3mm}
\caption{\textcolor{black}{Normalized inference energy consumption of (a) linear at TLC, attention at DRAM/pSLC with baseline as 3D-FPIM, (b) linear at TLC, attention at DRAM/pSLC with baseline as S-FLASH. BL: Baseline. AB: Activation buffering. DD: Distributed dataflow.}}
\vspace{-5mm}
\label{Fig_Energy}
\end{figure*}
\textbf{Workloads:}
\textcolor{black}{
To evaluate the latency and throughput of the NITRO architecture, we selected several widely used decoder-only Transformer models for benchmarks.
Specifically, we used LLaMa-2 \cite{touvron2023llama} with two parameter size types (7B and 13B) and OPT \cite{zhang2022opt} with four parameter size models (1.3B, 2.7B, 6.7B, and 13B).
This selection enables assessment of NITRO's performance under diverse computational demands.
In particular, it allows analysis of the relationship between various weight matrix dimensions and the physical resource constraints (e.g., number of rows or columns) of a NAND-PIM plane.
If the weight matrix dimensions exceed the capacity of an individual plane, partitioning to different WLs are required.
This can potentially increase computational iterations and influence overall system performance.
All model evaluations employed an 8-bit post-training quantization scheme.
However, the proposed MVM operation can be adapted to other weight precisions, such as INT4 or INT16, by appropriately modifying the mapping of weights across TLC cells and the number of ADC and PE.
Following the 3D-FPIM method \cite{lee20223d}, we also adopted analog current-sum method for MAC operations in NITRO's NAND-PIM.
We assume that the inherent NAND cell variations are comparable to those characterized in previous NAND flash study \cite{lee2020neuromorphic}, ensuring our computational accuracy remains at levels consistent with previous baselines.
Therefore, we omitted the accuracy evaluation in this paper.
}



\subsection{Performance}
\label{section_result_performance}

\textcolor{black}{
Fig. \ref{Fig_latency} shows the normalized inference latency of the proposed NITRO architecture compared against the 3D-FPIM (Fig. \ref{Fig_latency}a) and S-FLASH (Fig. \ref{Fig_latency}b) baselines. 
Using DRAM for activation buffering substantially reduces latency, with an average speedup of 4.07$\times$ against the 3D-FPIM baseline.
This improvement is due to the elimination of high-latency TLC NAND programming operations, which are required for buffering intermediate activations in the attention stages ($QK^T$, $SV$) and dominate baseline latency.
As shown in Fig. \ref{Fig_latency}a, using pSLC NAND for this purpose is less effective than DRAM.
While pSLC cell operations are faster, the page-wise multi-bit data layout (at least 7 pairs of BLs per byte) increases ADC multiplexing latency, canceling out the cell-level speed gain.
This implies that DRAM is more suitable for activation buffering.
}
Applying the distributed dataflow methodology provides further improvement in latency.
When combined with attention in DRAM-PIM scenario, the fully optimized NITRO architecture achieves final normalized latencies of approximately 0.003 and 0.069 relative to 3D-FPIM baseline and S-FLASH baseline, respectively.
The additional speedup is achieved from optimizing the linear block execution within TLC NAND-PIM through efficient weight mapping and exploiting multi-plane parallelism.
The improvement with distributed dataflow also benefits in the attention at pSLC configuration, reducing final normalized latencies approximately 0.29 and 0.448 against 3D-FPIM and S-FLASH, respectively.

Similar to latency reduction, the proposed NITRO architecture shows an improvement in inference throughput.
Fig. \ref{Fig_Throughput}a and Fig. \ref{Fig_Throughput}b show the normalized throughput compared to 3D-FPIM and S-FLASH, respectively. 
Enabling activation buffering, especially with DRAM, provides an increase in throughput by mitigating the read/program bottleneck and slow PIM computations in baseline approaches.
The fastest throughput is attained by full NITRO configuration.
NITRO shows a high concurrency by effectively parallelizing computations inside TLC NAND for linear blocks and leveraging optimized data distribution across multiple planes.
As a result, NITRO with attention at DRAM-PIM improves throughput by approximately 287x and 14.5x over 3D-FPIM baseline and S-FLASH baseline, respectively.
This throughput improvement highlights NITRO's effectiveness in exploiting available parallelism inside the NAND-PIM.

\subsection{Energy Consumption}
\label{section_result_energy}

\textcolor{black}{
We furthermore compared the inference energy consumption of NITRO with two baselines.
As shown in Fig. \ref{Fig_Energy}, the baseline energy consumption is dominated by the attention stages ($QK^T$, $SV$).
This is due to repeated energy-intensive read and program operations across the NAND cell array if the intermediate activations need to be programmed back to the TLC array.
NITRO's strategy of offloading attention blocks to low-power DRAM reduces the energy consumption thanks to the
faster read/write cycle and lower active power of the DRAM.
Combined with distributed dataflow for linear block computations in TLC array, the overall normalized energy consumption is reduced by approximately 98.9\% compared to the baseline (Fig. \ref{Fig_Energy}a).
When considering pSLC NAND for attention block execution instead of DRAM (Fig. \ref{Fig_Energy}a), the energy consumption shows a minor increase (less than 1\%) over the TLC-only baseline.
While pSLC offers faster read/program operations, its page-wise data layout for multi-bit values can increase concurrent utilization of ADCs and PEs.
This can increase peripheral energy, partially degrading pSLC gains.
However, the distributed dataflow ($BL+AB+DD$) optimizes peripheral utilization and reduces energy even in pSLC scenario.  
}

\textcolor{black}{
The characteristics of S-FLASH approach, such as limited accumulation groups and separate MSB/LSB weight placement, can cause weights to be distributed across more WLs.
This potentially increases MVM iterations, leading to an increase in energy consumption.
Despite this, NITRO's comprehensive optimization ($BL+AB+DD$) still achieves an average energy consumption reduction of approximately 72.6\% compared to the S-FLASH baseline (Fig. \ref{Fig_Energy}b).
In summary, NITRO architecture effectively reduces energy consumption using low-power DRAM for the attention layer and optimizing dataflow within NAND-PIM for the linear layer.
}
\newcolumntype{Y}{>{\centering\arraybackslash}X}

\begin{table}[t]
\begin{center}
\renewcommand{\arraystretch}{1.2}
\caption{Area breakdown}
\label{table_area}
\vspace{-2mm}
\textbf{NAND-PIM Plane Configuration}
\\[1mm]
\begin{tabularx}{\linewidth}{ Y | Y }
\Xhline{1pt}
Component & Area [mm\(^2\)] \\
\hline
Cell Array & 27.32  \\
ADC & 17.51 \\
Shift \& Accum. & 1.46 \\
Others & 0.37  \\
\hline
Total & 46.66 \\
\Xhline{1pt}
\end{tabularx}
\\[3mm]
\textbf{DRAM-PIM Bank Configuration \cite{shin2018mcdram}}
\\[1mm]
\begin{tabularx}{\linewidth}{ Y | Y }
\Xhline{1pt}
Component & Area [mm\(^2\)] \\
\hline
Cell Array$^{\mathrm{a}}$ & 5.22  \\
MAC Unit & 0.33 \\
Others & 1.53 \\
\hline
Total & 7.08 \\
\Xhline{1pt}
\end{tabularx}
\vspace{-2mm}
\end{center}
$^{\mathrm{a}}$\textcolor{black}{Assume area of cell array to be 60\% of total area \cite{chatterjee2017architecting}.}
\vspace{-3mm}
\end{table}


\subsection{Area}
\label{section_result_area}

Table \ref{table_area} shows the area breakdown of NAND-PIM and DRAM-PIM of the NITRO architecture.
To accurately digitize the potentially wide range of accumulated analog current sums resulting from MVM operations, 16-bit SAR-ADCs were modeled, based on previous work \cite{chung202116}.
To achieve a balance between area cost and digitization throughput, one ADC is allocated for every six BLs in TLC mode.
The area of \textit{Others} category, including decoders and load capacitors, is calculated using the 3D-FPIM simulator \cite{lee20223d}.
The area of DRAM-PIM was estimated based on the McDRAM architecture \cite{shin2018mcdram}.
Overall, NITRO requires a large area for high-resolution ADCs in the NAND flash peripheral circuit.
This tradeoff in area enables significant latency reductions by performing parallel computations directly inside the memory arrays and minimizing heavy data movements.


\section{Conclusion}
\label{section_conclusion}


We introduced NITRO, a high-performance NAND flash-based ISC architecture.
Conventional NAND-only PIM approaches face a performance bottleneck due to the slow latencies when handling intermediate data.
NITRO mitigates this by strategically offloading intermediate activation buffering and attention block computations to fast DRAM, while concurrently executing weight-intensive linear blocks efficiently inside the NAND-PIM using a distributed dataflow methodology that maximizes intra-plane and multi-plane parallelism.
Our evaluation results across various LLMs demonstrate that NITRO reduces inference latency by up to 99.7\% and approximately 93.1\% compared to the 3D-FPIM and S-FLASH baselines, respectively, while improving inference throughput by approximately 287× and 14.5×, respectively.

\appendices
\section{Major Extensions over the Conference Version}
\label{appendices}

Compared to the conference version accepted for publication in DATE 2026, several major updates and extensions have been added to the extended submission:

\begin{enumerate}
    \item Compared to the conference version, this extended paper includes a detailed background on 3D NAND flash memory. Specifically, Fig. \ref{prelim_3D_NAND_Arch} was added to illustrate the die-level and plane-level architectures and the voltage distribution of triple-level cells (TLC), and Section \ref{section_prelim_3d_nand} provides an in-depth explanation of cell organization, the bitline/wordline structure, and the multi-bit read sequence using reference voltages $V_{r0}$--$V_{r6}$.

    \item Compared to the conference version, this extended paper provides a detailed explanation of the 3D NAND flash-based matrix-vector multiplication (MVM) mechanism. Specifically, Fig. \ref{prelim_3D_NAND_MVM} was added, and Section \ref{section_prelim_MVM} describes how weights are stored in cells, how activations are encoded as bit-wise voltage pulses, and how the analog current-sum method with paired bitlines and ADCs realizes signed MVM operations.

    \item Compared to the conference version, this extended paper introduces a dedicated Overview subsection for the proposed architecture. Specifically, Fig. \ref{overview_arch} was added, and Section \ref{section_proposed_overview} contrasts the baseline NAND-PIM with NITRO at the system level and motivates both the activation DRAM buffering and the distributed dataflow strategies.

    \item Compared to the conference version, this extended paper provides a more detailed explanation of the activation DRAM buffering scheme. Specifically, Figs. \ref{contribution_1} and \ref{contribution_1_dataflow} were added, and Section \ref{section_proposed_AB} elaborates how intermediate activations such as $K$ matrices are offloaded to DRAM-PIM to avoid TLC NAND programming stalls, and how Transformer layers are mapped to TLC NAND-PIM (linear blocks) and DRAM-PIM (attention blocks). The pSLC alternative for DRAM-less SSD configurations is further elaborated in this extended version.

    \item Compared to the conference version, this extended paper further expands the distributed dataflow discussion. Specifically, Fig. \ref{contribution_2} was added, and Section \ref{section_proposed_DD} separates the description into Linear Blocks and Attention Blocks, elaborating weight/activation partitioning, multi-plane parallelism, DRAM-bank-level head distribution, and pSLC diagonal placement.

    \item Compared to the conference version, this extended paper provides a comprehensive description of the NITRO hardware architecture. Specifically, Fig. \ref{NITRO_Architecture} was added, and Section \ref{section_proposed_NITRO_Arch} details the top-level NITRO SSD architecture including the BLS decoder, ADCs with integrators, processing elements (PE) for digital accumulation, and the bypass path for conventional NAND read/program operations. The 8-bit weight computation flow with sign-and-magnitude representation across multiple physical NAND cells is also explained.

    \item Compared to the conference version, this extended paper includes additional experimental evaluations. Specifically, Fig. \ref{Fig_Throughput} (inference throughput) was added to report normalized inference throughput in addition to the latency and energy results. Moreover, S-FLASH \cite{kang2021s} was newly included as a second baseline alongside 3D-FPIM \cite{lee20223d}, enabling a broader comparison with representative NAND flash-based PIM approaches. The LLaMa-2 (7B/13B) and OPT (1.3B/2.7B/6.7B/13B) workloads used in the conference version were retained and further analyzed with the added throughput metric and the additional S-FLASH baseline, as discussed in Section \ref{section_result}.

    \item Compared to the conference version, this extended paper expands the brief area discussion into a dedicated Area subsection. Specifically, Table \ref{table_area} was added, and Section \ref{section_result_area} provides component-level area breakdowns for NAND-PIM and DRAM-PIM, clarifying the cost and allocation of 16-bit SAR-ADCs and the DRAM-PIM area estimation.
\end{enumerate}

\bibliographystyle{IEEEtran}
\bibliography{bstcontrol,bibliography}

\end{document}